\documentclass[12pt]{article}

\usepackage{wrapfig}
\usepackage{graphicx}
\begin{document}
\begin{flushright}
January 2003
\end{flushright}
\vspace{3mm}
\begin{center}
{\bf\Large Counting Extra Dimensions: Magnetic Cherenkov Radiation from High Energy
Neutrinos}\\[3mm]
G.~Domokos$^{a}\footnote{e-mail: skd@jhu.edu}$, Andrea Erdas$^{a,b}$\footnote{e-mail: aerdas@pha.jhu.edu} 
and S.~Kovesi-Domokos$^{a 1}$\\
$^{a}$ Department of Physics and Astronomy, The Johns Hopkins University, Baltimore, MD
\\
$^{b}$ Dipartimento di Fisica dell'Universit\'{a} \& INFN, Cagliari
\end{center} \vspace{3mm}
{\small {\em Abstract} In theories which require a space of dimension $d >4$, there is a 
natural mechanism of suppressing neutrino masses: while Standard Model fields are 
confined to a 3-brane, right handed neutrinos live in the bulk.
 Due to Kaluza-Klein excitations, the effective magnetic moments of neutrinos are
 enhanced. The effective magnetic moment is a 
monotonically growing function of the energy of the neutrino:
 consequently, high energy
neutrinos can emit observable amounts of magnetic Cherenkov radiation. By observing the 
energy dependence of the magnetic Cherenkov radiation, one may be able to determine
the number of compactified dimensions.
\vspace{5mm}
\section{Introduction}
In theories with extra dimensions there is an interesting alternative to the 
conventional seesaw mechanism of giving neutrinos a small mass~\cite{nima}.
In that scheme, the left-handed neutrino ($\nu_L$) resides on the brane together with all
other Standard Model fields, whereas the right handed neutrinos ($\nu_R$) live in the bulk.
The neutrino mass of any flavor is then suppressed with respect to other fermion masses 
by a factor of the order of $\sim M_{\ast}/M_P$, 
$M_P$ being the four dimensional Planck mass and $M_{\ast}$ is the string scale. Its  currently favored
 value is somewhere between $\sim 10$\,TeV  and $\sim 100$\,TeV. Phenomenological implications of such a scheme
 have been
studied by a number of authors. For our purposes, the relevant papers are the ones authored by
 Faraggi and Pospelov~\cite{Faraggi}, McLaughlin and Ng~\cite{ng} and by  Agashe and Wu~\cite{agashe}.

We argue that if indeed $\nu_{R}$ of any flavor lives in the bulk, then magnetic Cherenkov radiation
by that neutrino flavor is enhanced due to the contribution of Kaluza-Klein (KK) excitations 
to the transition probability.
In the next Section we outline the calculation leading to such a conclusion.  
We find that, in fact,  there is hardly anything to calculate, since one is able to use results
of ref.~\cite{ng} combined with those of Sawyer~\cite{sawyer}. In Sec.~3 practical aspects of the
results are discussed; in particular, we estimate the yield of Cherenkov photons in water,
relevant for underwater detectors.(Estimates of yields in detectors in ice
are somewhat lower, but not dramatically so.) 
  The last Section is devoted to a discussion of the results.

It is to be noted that there exists an extensive literature on magnetic Cherenkov
 radiation of neutrinos:
several results have been repeatedly rediscovered, rederived, {\em etc.} often with results which 
contradict each other. To our knowledge, the papers of Sawyer~\cite{sawyer} and of Ionnisian and
Raffelt~\cite{ion} are internally consistent. In addition, ref.~\cite{ion} contains an extensive list of references. 
Throughout this paper natural units are used:
$\hbar = c = 1$.
\section{Cherenkov radiation in the presence of KK excitations}
If a particle is neutral, but of nonzero spin,
it is scattered on an electromagnetic field due to its magnetic moment. By using the optical
theorem and crossing symmetry, the magnitude of the magnetic moment can be extracted from the 
imaginary part of the  forward (magnetic) Compton amplitude in the zero frequency limit. 
If the particle in
question has excited states, this leads to an energy dependent {\em effective magnetic moment}:
the imaginary part of the Compton amplitude depends on the number of excited states available at
a given energy. This is, in fact, the argument put forward in ref.~\cite{ng}.

The forward Compton amplitude depends on the expression of the transition moments between the 
ground state of the particle and its excited states. Obviously, this reasoning is
applicable to neutrinos with $\nu_R$ living in the bulk, due to
the existence of KK excitations. 
  Assuming  toroidal compactification 
of the extra dimensions of equal radii ($R^{-1}=M_{c}$), it is obvious that the effective magnetic moment squared
is proportional to the number of available KK states at a given energy, as well as to the mass squared 
 of the neutrino:
\begin{equation}
\mu_{\rm eff}^{2} \propto m_{D}^{2}\left( \frac{E}{M_{c}}\right)^{n}
\end{equation}

 McLaughlin and Ng~\cite{ng}
find for a reasonable choice of parameters, qualitatively consistent with solar and 
atmospheric neutrino data:
\begin{equation}
\mu_{N} \approx 1.6\times 10^{-19}\mu_{B} \frac{m_{D}}{1 {\rm eV}},
\end{equation}
where $\mu_{N}$ is the transition moment between the zero mode of $\nu_{L}$ and  the $N^{\rm th}$
KK level, while $m_{D}$ stands for   the (Dirac) mass of the neutrino. 
The quantity $\mu_{B}$ is the Bohr magneton,  giving  a 
convenient comparison scale.  The result quoted above is valid in the weak mixing approximation, 
which is well justifiable
here, see~\cite{Faraggi, ng}. Remarkably, the transition moment is independent of the KK
level: this fact plays a significant role in what follows.

The resulting effective magnetic moment  is given by:
\begin{equation}
\mu_{\rm eff}^{2}\approx 1.6\times 10^{-15} \mu_{B}^{2}
 \frac{\pi^{(1-n)/2}}{2^{n}\Gamma \left( \frac{n+1}{2}\right)}
\left(\frac{E}{M_{\ast}}\right)^{n},
\end{equation}
where $n$ stands for the number of compactified dimensions. The factor multiplying 
$(E/M_{\ast})^{n}$ is proportional to the degeneracy factor of the KK states.
In the last equation, the mass of the neutrino has been eliminated in favor of its Yukawa coupling to the Higgs
field, $y$, according to the formula:
\begin{equation}
m_{D}^{2}= y^{2}V^{2}\left(\frac{M_{c}}{2\pi M_{\ast}}\right)^{n},
\end{equation}
see ref.~\cite{nima}.
Here, $V$ stands for the vacuum expectation value of the Higgs field; its standard value is being used. 
Due to our ignorance, we also  set $y\approx 1$.
 As far as  $n$, the number of compactified dimensions  is concerned, certain models
(superstring, SUGRA) suggest $n=6$ or $n=7$, respectively. However, it should be realized that
at present, it is unclear whether we have a phenomenologically viable model describing physics beyond
the Standard Model. For that reason, we prefer to keep an open mind about the number of compactified 
dimensions.

With present techniques,  Cherenkov radiation can be detected  typically in the visible and in 
the (near) UV.
Hence, one can directly use the emission rates obtained by Sawyer~\cite{sawyer}, replacing 
$\mu^{2} \longrightarrow \mu_{\rm eff}^{2}$. (This approximation is justified by the fact
that the energy  of the emitted Cherenkov photon is low compared to energy scales of the KK
excitations.)

Using this, one obtains the formula for the number of photons emitted in a unit frequency interval
and per unit length by a neutrino of $v\approx 1$:
\begin{equation}
\frac{dN}{d\omega dx}= \omega^{2}\left( \mu_{\rm eff} \right)^{2} \left( \epsilon (\omega) -1\right).
\label{emissionrate}
\end{equation}

It is amusing to observe that in the tree approximation used by Sawyer~\cite{sawyer}, the result of a 
lowest order perturbative calculation of the Cherenkov emission rate gives the same result as
the old calculation of Ginzburg based on classical electrodynamics~\cite{ginzburg}. 
This is a fairly general result: the tree
approximation to any given Feynman diagram can be expressed in terms of an iterative solution of the
classical field equations to the given order. Taking into account KK excitations requires the use
of a quantum calculation of the type devised by Sawyer, {\em l.c.} However, in the {\em effective
moment approximation} as used above, one can still resort to the classical calculation.
\section{Observability}
Without the enhancement provided by the presence of KK excitations, it is hopeless to observe the 
magnetic Cherenkov radiation. Using the upper limits on the magnetic moments of neutrinos as given 
by the Particle Data Group~\cite{pdg}, one finds that the number of photons emitted by a neutrino
of any flavor is $\leq 10^{- 16}$/km or so. Clearly, the observation of such a rare event, let alone its
identification as a Cherenkov emission is out of question in any detector at present
 or in the foreseeable future.

{\em Do KK excitations help?}

In order to answer that question, we assume a quantum efficiency of the PMTs in a detector
to be around 10\% and that about 100 photoelectrons have to be detected in a detector in
order to identify a Cherenkov ring. Furthermore, we assume a detector size of the order of 1km.
A simple model of phototube response consists of no response below a minimal wavelength, 
$\lambda_{min}$,  in the UV, a flat response between $\lambda_{min}$ and a fixed $\lambda_{max}$,
taken to be $\lambda_{max}\approx 0.55\mu$m and no response again for $\lambda > \lambda_{max}$.
This appears to be a reasonable zeroth approximation to phototube response~\cite{nemethy}
leading to the requirement that about 1,000 photons have to be emitted for 
$\lambda_{min}< \lambda < \lambda_{max}$.
We integrate eq.~(\ref{emissionrate}) for a fixed  $\lambda_{max}=550 \mu$m and for several 
values of $\lambda_{min}$, using an empirical formula~\cite{nemethy} for the frequency dependence of the
index of refraction of water. This formula is a  fair approximation to the frequency 
dependence found in
ref.~\cite{freqdependence}. The empirical formula reads:
\begin{equation}
\epsilon = 1.76253 -0.0133998 \left(\lambda \right)^2
+ \frac{ 0.00630957}{ \left(\lambda \right)^{2} - 0.0158806},
\label{eq:empirical}
\end{equation}
the wavelength $\lambda $ being measured in $\mu$m.
Using (\ref{eq:empirical}) one has to calculate the integral:
\begin{equation}
F= \int_{\lambda_{min}}^{\lambda_{max}}d\lambda \frac{1}{\epsilon \lambda^{2}}
\left( \epsilon -1 \right) \frac{d}{d\lambda}\frac{ 1}{\sqrt{\epsilon} \lambda}
\label{eq:frequencyintegral}
\end{equation}
The integration over frequencies was replaced by an integration over wavelengths,
using the relation $\omega \lambda = 2\pi  \epsilon^{-1/2}$. 
The integral $F$ is plotted as a function of $\lambda_{min}$ in  Fig.~\ref{Fintegral},
while keeping $\lambda_{max}$ fixed.
\begin{figure}[h]
\includegraphics[width=0.8\textwidth]{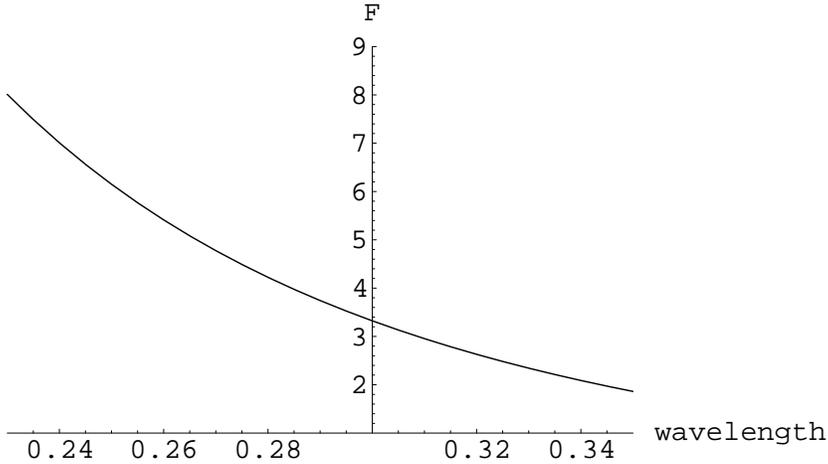}
\caption{The function F in eq.~\ref{eq:frequencyintegral} plotted as a function of
$\lambda_{min}$, keeping $\lambda_{max}$ at 550$\mu$m. Units: F is plotted 
 in $\mu{\rm m}^{-3}$ and $\lambda$ in $\mu{\rm m}$.}
\label{Fintegral}
\end{figure}
It is to be noted that in contrast to Cherenkov radiation by a charge, the emission rate 
grows as $\omega^{2}$, thus magnetic Cherenkov radiation is more sensitive to detector response
in the ultraviolet.

In the following table we list values of the  energy, $E_{1000}$, necessary to generate 
1,000 Cherenkov photons over a distance of 1~km,  (typical of the size of future 
detectors) for some relevant numbers of compactified dimensions. 
Because of the structure of eq.~(\ref{emissionrate}), the  energies 
listed are measured in units of $M_{\ast}$.
\begin{table}[h]
\caption{$E_{1000}/M_{\ast}$ needed  for the  emission of 1,000 Cherenkov photons/km}
\begin{center}
\begin{tabular}{||l|c|c|c|c|c|c|c|c|c|c||}\hline
n & 2 & 3 & 4 & 5 & 6 & 7 & 10 & 14 & 17 & 20\\ \hline
$^{10}\log (E_{1000}/M_{\ast})$ & 10 & 6.88 & 5.33 & 4.41 & 3.8 & 3.37 & 2.62 & 2.14 & 1.99 & 1.84\\ \hline
\end{tabular}
\end{center}
\end{table}
It is to be noted that the table above can only serve as a benchmark. In any present and future detector, 
the minimal number of photons required depends on a geometrical factor: the fraction of the 
detector volume or surface covered by PMTs, depending on the specific detector type. 
Since the coverage is never 100\%, the minimal number of photons is higher than exhibited above.
However, based on the data presented, it is easily calculable for any specific detector.

So far, we have not taken into account an essential feature of the theory as described here. {\em In 
order for the usual  theory of Cherenkov radiation be valid, one has to make sure that the 
target medium responds collectively and not as a collection of individual atoms.} One can establish
a criterion for a collective response rather easily. 

Assuming that the medium responds as a macroscopic body, one finds that it can take up momentum,
but not energy. As a consequence, the equation of energy conservation reads:
\begin{equation}
p_{i}\approx E_{i} = \sqrt{p_{f}^{2} + M_{KK}^{2}} + \omega .
\label{eq:energyconservation}
\end{equation}
In eq.~(\ref{eq:energyconservation}), $\omega$ stands for the energy of the emitted photon, 
$p_{i}, p_{f}$ for the initial and final momenta of the neutrino, respectively and $M_{KK}$
for the mass of the KK excitation. On noting that $\omega$ is typically in the visible or in the
near-UV, one finds that its contribution  can be safely neglected in eq.~(\ref{eq:energyconservation}).
 Thus, from  eq.~({\ref{eq:energyconservation}), one easily finds an 
approximate expression of the momentum transferred to the target medium, {\em viz}
\begin{equation}
q = p_{i} - p_{f} \approx \frac{M_{KK}^{2}}{2E_{i}}
\label{eq:momentumtransfer}
\end{equation}
The medium responds collectively if $1/q \geq d$, where $d$ is the average
 intermolecular distance. In water $d\approx 10^{-8}$~cm; the corresponding
energy being $1/d \approx 2$~keV. Very little is known about water or ice under
extreme conditions, such as the ones prevailing at the sites of NESTOR or AMANDA.
It is expected that $d$ is smaller there. However, we use the above value to be on the
safe side. In order to obtain an estimate of the neutrino energies needed, we take
$N=1$, so that $M_{KK}\approx M_{c}$. It follows that the initial energy of an incident neutrino
has to satisfy the approximate inequality:
\begin{equation}
\frac{E_{i}}{M_{c}}\geq \frac{M_{c}d}{2}
\label{eq:coherence}
\end{equation}
\section{Discussion}
It is clear from   Table~1 that low numbers of extra dimensions are disfavored from
the point of view of the observability of magnetic Cherenkov radiation.
 In view of the stringent lower limits on
$M_{c}$ given by Hannestad and Raffelt~\cite{raffelt} it is unreasonable to expect any effect
even at $E\approx 10^{20}$~eV. The situation is somewhat unclear for values of $n$ motivated by
currently existing theoretical schemes, ($n=6,7$), since currently there are no stringent 
limits  comparable  to the ones given by ref.~\cite{raffelt} for a low number of
compactified dimensions. Future limits on $n$ obtained from other sources will determine whether or not 
magnetic Cherenkov radiation is a useful tool in the search for large compactified extra dimensions.

For now, we  take  the limits obtained by Anchordoqui~{\em et al.}, ref.~(\cite{Anchordoq}) as a guide, 
as well as our recent estimate of the value of $M_{\ast}$ \cite{amsterdam}. Anchordoqui~{\em et al.}
give lower limits on $M_{\ast}$ about 1~or~2~TeV, whereas, in ref.~\cite{amsterdam} 
we estimate $M_{\ast}\approx 80$~TeV. The requirement of coherent response, eq.~\ref{eq:coherence}
depends explicitly on $M_{c}$. Given symmetric toroidal compactification, as assumed, one can calculate $M_{c}$
from $M_{\ast}$ using the ADD formula, \cite{ADD}:
\begin{equation}
M_{c} = 2\pi M_{\ast} \left( \frac{M_{\ast}}{M_{P}}\right)^{2/n}.
\end{equation}
 Using this information, we list the minimal 
incident neutrino energy (in the laboratory system) assuming $n=6$~as required by a supersting
for the values of $M_{\ast}$ mentioned above.
\begin{table}[h]
\begin{center}
\caption{Values of the compactification mass and minimal value of of $E_{\rm i}$   for $n=6$, typical values of $M_{\ast}$}
\begin{tabular}{|c|c|c|}
\hline
$M_{\ast}$ in TeV& $M_{c}$ in GeV &$E_{\rm i}$ in TeV\\ \hline
1 & 0.028 & 0.12\\ \hline
2 & 0.069 & 1.2 \\ \hline
10& 0.6 &  90 \\ \hline
80 & 9.4 & 22,000\\
\hline
\end{tabular}
\end{center}
\end{table}

Clearly, very high neutrino energies are needed in order to make magnetic Cherenkov 
radiation emitted by neutrinos of any flavor  observable. As a consequence, the coherence requirement 
is not a very serious issue.
However, due to its unique features, such a radiation can serve as a very useful diagnostic
tool in future detectors.

{\em Acknowledgements}. We thank Peter~Nemethy for useful discussions on the criteria of observability
of Cherenkov radiation and for supplying the empirical formula for the index of refraction. Shmuel Nussinov
kindly drew our attention to the limitations of the collective description of a continuum.
We thank Haim Goldberg for pointing out an error in the original version
of this paper. A.E. thanks the
{\em Ministero dell'Istruzione, dell'Universit\'{a} e della Ricerca} for partial support of this research
under {\em Cofinanziamento P.R.I.N. 2001}.
 G.D. and S.K.D. thank G.~Pint\'{e}r for first raising the possibility of using  magnetic
Cherenkov radiation as a tool in the search for ``new physics''.

\end{document}